\documentstyle[prl,preprint,aps]{revtex}
\begin{document}
 
\draft
\title{Universal features of the off-equilibrium 
fragmentation with the Gaussian dissipation}

\author
{Robert Botet$^{\dagger}$ and  Marek 
P{\l}oszajczak$^{\ddagger}$}
 
\address{$^{\dagger}$
 Laboratoire de Physique des Solides,
B\^{a}timent 510, Universit\'{e} Paris-Sud
\\ Centre d'Orsay, F-91405 Orsay, France  \\
and   \\  $^{\ddagger}$
Grand Acc\'{e}l\'{e}rateur National d'Ions Lourds 
(GANIL), \\
CEA/DSM -- CNRS/IN2P3, BP 5027,  F-14076 Caen Cedex 05, 
France  }
 
\date{\today}
 
\maketitle
 
\begin{abstract}
\parbox{14cm}{\rm
 
We investigate universal features of the 
off-equilibrium sequential 
and conservative fragmentation processes with 
the dissipative effects  which are simulated
by the Gaussian random inactivation process.
The relation between the fragment multiplicity scaling 
law and the fragment
size distribution is studied and a dependence of 
scaling exponents on the
parameters of fragmentation and inactivation rate 
functions is established.
\smallskip\\}
\end{abstract}

\vfill
\newpage
 
\narrowtext

Fragmentation is the universal process which 
can be found at all scales in the nature. The most 
general sequential binary 
and conservative fragmentation processes with
scale-invariant fragmentation and inactivation rate 
functions, 
have been previously studied in much 
details\cite{sing1,sing0,kno}~. 
The phase diagramme of these off-equilibrium processes 
has been established 
and the universal aspects of both the fragment size
distribution and the total number of 
fragments distribution (i.e., the 
multiplicity distribution) have been 
determined\cite{kno,prln}~. 
In this fragmentation-inactivation binary (FIB) 
model\cite{sing1,sing0}~,  
one deals with fragments characterized by some conserved 
scalar
quantity that is called the fragment mass~. The 
anscestor fragment of mass $N$~
is fragmenting via an ordered and irreversible sequence 
of steps.
The first step is either a binary fragmentation,
$(N) \rightarrow (j)~+~(N-j)$~, or an
inactivation $(N) \rightarrow (N)^{*}$~.
Once inactive, the cluster cannot be reactivated 
anymore. The fragmentation 
leads to two fragments, with the mass partition 
probability $\sim F_{j,N-j}$~.
In the following steps, the 
process continues independently for each active 
descendant fragment until
either the low mass cutoff for further indivisible 
particles
(monomers) is reached or all fragments are inactive.
For any event, the fragmentation
and inactivation occur with the probabilities per unit 
of time
$\sim F_{j,k-j}$~ and $\sim I_k$~ respectively.
The fragmenting system  and  its  evolution is 
completely specified by that
rate functions and the initial state. It is also useful 
to consider
the fragmentation probability $p_F$~
without specifying masses of descendants :
$p_F(k) = \sum_{i=1}^{k-1} F_{i,k-i}~( I_k +
\sum_{i=1}^{k-1} F_{i,k-i} )^{-1}~$~ . 
If the instability of smaller fragments is smaller than 
instability of larger 
fragments, $p_F(k)$~ is an increasing  function  of 
fragment mass and the total mass is converted into 
finite size fragments. This is the shattered phase. The fragment mass 
independence of $p_F(k)$~ at any stage of the process until the 
cutoff-scale for monomers characterizes the critical 
transition region. The
multiplicity anomalous dimension : $\gamma = d(\ln 
<m>)/d(\ln N)$~,  is
the order parameter in the FIB model. It equals 1 in the 
shattering phase 
and takes the intermediate value between 0 and 1 
in the critical transition region.

For most fragmenting systems, the off-equilibrium 
relaxation process ceases 
due to a dissipation. The dissipation is not always 
scale-invariant as 
considered in Ref. 1 but, on the contrary, it is often 
characterized by a 
definite and usually small length scale. It is then an 
open question to which extent the fragmentation 
processes 
which on one side are 
driven by the homogeneous scale-invariant fragmentation 
rate function and on other 
side are inactivated at a certain fixed scale by the 
random 
inactivation process, may develop scale-invariant and 
universal features in both the fragment mass 
distribution $n(k)$~ and the fragment multiplicity
distribution $P(m)$~. This question is important
in view of the widespread occurence of scale-invariant 
fragment mass distributions $n(k) \sim k^{-\tau}$~ 
and the lack of convincing arguments for
using homogeneous dissipation functions in many 
processes including
parton cascading in the perturbative quantum 
chromodynamics 
(PQCD)\cite{future}~ or
the fragmentation of highly excited atomic nuclei, 
atomic clusters or polymers. In this work, we address this fundamental 
question using the 
FIB process with the homogeneous fragmentation rate 
function : 
$F_{j,k-j} = [j(k-j)]^{\alpha}$~, and with the
dissipation at small scales which is modelled by the
Gaussian inactivation rate function : 
\begin{eqnarray}
\label{inact}
I_k = c\exp 
[-\frac{1}{2{\sigma}^{2}}\left( \frac{k-1}{N} 
\right)^{2}]~~~\ .
\end{eqnarray}

An asymptotic ($t \rightarrow \infty$) fragment mass 
distribution in the
critical transition region of FIB 
model with scale-invariant dissipation 
phenomena\cite{sing1,sing0}~, 
is a power law with an exponent $\tau \leq 2$~. In the
shattering phase, the fragment mass distribution is also 
power law but with an
exponent $\tau > 2$~. Another characteristic observable 
is the fragment multiplicity distribution : 
$P(m) = \sum_k^{} P_k(m)$~,
where $P_k(m)$~ is the probability distribution of the 
number of fragments of mass $k$~. This quantity has been 
intensely 
studied in the strong interaction physics\cite{carr}~. 
Of 
particular importance 
is a possibility of asymptotic scaling of 
multiplicity probability distributions :
\begin{eqnarray}
\label{b4}
<m>^{\delta} P(m) = \Phi (z_{({\delta})})~~,~~~~~~~~
z_{({\delta})} \equiv \frac{m-<m>}{<m>^{\delta}}~
\end{eqnarray}
where the asymptotic behaviour is defined as
$<m> \rightarrow \infty$~, $m \rightarrow \infty$~ for a 
fixed \\
$(m/<m>)$ -- ratio. $<m>$~ is the multiplicity of 
fragments averaged over an ensemble of events. The 
scaling law 
(\ref{b4}) means that for example data for differing 
energies
(hence differing $<m>$) should fall on the same curve 
when
$<m>^{\delta} P(m)$~ is plotted against the scaled 
variable
$z_{({\delta})} \equiv (m - <m>)/<m>^{\delta}$~.
Some time ago Koba, Nielsen and Olesen 
(KNO) suggested an asymptotic scaling 
(\ref{b4}) with $\delta = 1$~ in the strong interaction 
physics\cite{koba}~. The same
scaling has been found also in the critical transition 
region of 
scale-invariant FIB process for $p_F>1/2$~ and $\alpha 
\geq -1$~\cite{kno}. 
Recently, Botet, P{\l}oszajczak and Latora
(BPL) reported another scaling limit in (\ref{b4}) with 
$\delta = 1/2$~, which holds  in the
percolation and in the shattering phase of 
scale-invariant 
FIB process\cite{prln}~. $\delta = 1/2$ and 1 are the 
two limiting values since 
$\delta > 1$~ or $\delta < 1/2$~ are incompatible with 
the scaling hypothesis (\ref{b4})~. 

The study presented in this Letter correspond to the 
domain $\alpha \geq -1$~ of fragmentation rate functions 
$F_{j,k-j}$~. Many known
homogeneous fragmentation kernels correpond to this 
domain. These include the 
singular kernel $\alpha = -1$~ in the PQCD 
gluodynamics\cite{doksh}~
, $\alpha = -2/3$~ for the spinodal volume instabilities 
in three dimensions\cite{sing0}~, $\alpha = +1$~ in the 
scalar 
$\lambda {\phi}_{6}^{3}$~ field theory in six 
dimensions\cite{scalar}~, and many others\cite{sing0}~.
 For $\alpha < -1$~, the fragmentation process is 
dominated by the
splitting  $(k) \rightarrow (k-1) + (1)$~ at each step 
in the cascade, and leads to the finite limiting value 
of $<m>$~ independently of the initial size 
$N$~\cite{kno}~. 
In this evaporation phase, the scaling solution 
(\ref{b4}) does not hold
and the multiplicity anomalous dimension  
is equal zero when $N \rightarrow \infty$~. This phase 
is not relevant 
for the problem we want to address in this
Letter. 

Without restricting the generality of our discussion, 
we will present below results for fragmentation kernels 
with : $\alpha = -1$~ and $\alpha = +1$~.
The upper part of Fig. 1 shows multiplicity 
distributions for $\alpha =
-1$~ in the scaling variables (\ref{b4}) for $\delta = 
1$~ (the upper left part), and fragment mass 
distributions 
for the same parameters (the upper right part). The 
cascade equations of Gaussian FIB model have been 
solved by Monte-Carlo simulations \cite{sing1,sing0}~ 
for 
different initial system sizes ($N = 1024, 4096$)  
and for the following exemplaric parameters :
$c = 1$~ and $\sigma = 0.1$~, $1$~ of inactivation rate 
function 
$I_k \equiv I_k(c,\sigma)$~. We have made exhaustive
analysis of $P(m)$~ for a broad range of $c, \sigma$~ 
parameters , finding in all cases the KNO scaling ($\delta = 1$).
We have found the KNO scaling  
{\it uniquely} for $\alpha = -1$~.
The shape of KNO scaling function 
${\Phi}(z_{(1)})$~ depends on the precise value of both 
$c$~ and $\sigma$~.

In the lower left part of Fig. 1, we show typical
multiplicity distributions for $\alpha = +1$~ which are 
plotted for different system sizes in
the BPL scaling variables ($\delta = 1/2$~).
The corresponding fragment mass distributions are shown 
in the lower right part
of Fig. 1. Again, the precise form
of BPL scaling function ${\Phi}(z_{(1/2)})$~ depends 
on the chosen set of parameters $c$~ and $\sigma$~. 
In contrast to these results of Gaussian FIB model, 
fragmentation process in the scale-invariant FIB 
model for any value of exponent $\alpha$~ may be found 
either in the 
critical transition region 
or in the shattering phase depending on the homogeneity 
index $\beta$~ of the inactivation rate function $I_k = 
I_1 k^{\beta}$~
\cite{sing1,sing0}~. 
This means that e.g. both for $\alpha = -1$~ and $+1$, one may see 
either the KNO scaling or the BPL scaling of multiplicity 
distributions depending on
the precise value of the homogeneity index of the 
inactivation term.

Concerning the fragment mass distributions, Fig. 1 shows
the distributions for $\alpha = -1, +1$~ and different 
parameters 
of Gaussian inactivation rate function $I_k(c,\sigma)$~. 
For $\sigma$ larger than $\sim 0.5$~, 
one finds the power law distribution of fragment masses
for any value of parameter $c$~. In the studied case :
$\sigma = 1$, $c = 1$~, the exponent
$\tau$~ equals 1.8 and 2.8 for $\alpha = -1$~ and 
$\alpha = +1$~ 
respectively. For a given $\alpha$~, 
the value of exponent $\tau$~ is remarkably independent 
of
$\sigma$~ but depends strongly on
the value of parameter $c$~ in $I_k(c,\sigma)$~. For a 
smaller value of
$\sigma $~ ($\sigma = 0.1$~ is shown in Fig. 1), the 
fragment mass distribution decreases exponentially 
and the shape of scaling function resembles the Gaussian 
distribution. The form of
this exponential distribution depends both on $c$~ and 
$\sigma$~ parameters.

As a generic case
for $\alpha = -1$~, we have found the scale-invariant 
region of power law fragment mass distributions with $\tau \leq 
2$~ for $\sigma$~ above $\sim 0.5$~, and 
the exponential region of mass distributions for $\sigma$~ less than $\sim 0.5$~
. The power law region
is completely analogous to the critical transition 
region of scale-invariant FIB model for $\alpha > -1$~ and 
$p_F > 1/2$~\cite{sing1,sing0,kno}~,
because the multiplicity anomalous dimension in both 
models is : 
\begin{eqnarray}
\label{tau}
\gamma = \tau - 1 ~~~~~~~( 0 \leq \gamma \leq 1 ) \ .
\end{eqnarray} 
We have verified validity of this relation in Gaussian FIB model for a broad 
range of $c, \sigma$~ values. In the exponential region
, $\gamma$~
is {\it always} equal 1 independently of the value of 
parameter
$c$~, i.e. this region is in the shattering 
phase. One should remind that shattering 
in the scale-invariant FIB model is related exclusively 
with the BPL scaling 
, whereas in the Gaussian FIB model for $\alpha = -1$~ 
the KNO 
scaling holds.

The fragment size distributions for $\alpha = +1$~ and 
different values of $\sigma$~ behave similarly as for 
the $\alpha = -1$ case, except that now for $\sigma $ above $\sim 0.5$~
the power law 
exponent is $\tau > 2$~. For all 
$\sigma $, i.e. in both exponential and power law 
regions of mass distribution, 
the multiplicity anomalous dimension 
is $\gamma = 1$~  and the BPL 
scaling holds.
This generic situation
is completely analogous to the multiplicity behaviour found in the shattering 
phase of scale-invariant FIB model\cite{sing1,sing0}~.

Whenever the fragment size distribution is a power law, 
the
KNO scaling of multiplicity distributions is associated 
with $\tau \leq 2$~ and 
the BPL scaling of multiplicity distributions
with $\tau > 2$~ in both scale-invariant and 
scale-dependent regimes of dissipation. This clearly 
indicates a direct
relation between the multiplicity scaling law and the 
fragment mass distribution scaling regimes in the FIB 
model. 
In view of the generality of FIB process, it
would be very interesting to test this relation 
experimentally. 
A novel aspect of the Gaussian FIB
model is associated with properties of multiplicity 
scaling in the new region of exponential fragment mass 
distributions. In this region, BPL scaling holds for $\alpha=+1$
whereas KNO scaling is seen for $\alpha = -1$~.

In Fig. 2 we plot for different values of the parameter $c$~  
the normalized cumulant factorial moment of order two\cite{foot} :
${\gamma}_2 = (<m(m-1)>-<m>^2)/<m>^2$~, 
vs the width $\sigma$~ of inactivation rate function 
$I_k(c,\sigma)$~.
The exponent of homogeneous fragmentation kernel is :
$\alpha = -1$~. For this choice of $\alpha$~, 
the KNO scaling holds and ${\gamma}_2$~ becomes the second moment
of scaling function ${\Phi}(z_{(1)})$~ which is independent of initial mass
$N$\cite{carr,kno}~. 
For each point ($c$~, $\sigma$)~, cascade equations of the 
FIB model have been solved exactly by recurrent formula 
\cite{kno} 
up to initial system size $N = 2^{18}$~.
As can be seen in Fig. 2, the
multiplicity fluctuations as measured by ${\gamma}_2$~ 
are extremely small in
the exponential region for $\sigma$~ less than  
$\sim 0.5$~. The change 
of ${\gamma}_2$~ when passing from the
power law to exponential region is continuous but the 
largest variations of
${\gamma}_2(\sigma)$~ appear at $\sigma \sim 0.5$~. 
For large values of $\sigma$~,
the cumulant factorial moment approaches a limiting 
value which depends on the
value of parameter $c$~.

The experimental informations about ${\gamma}_2$~ are not 
numerous 
and concern mainly charged particle multiplicities at relativistic 
and ultrarelativistic
energies. The DELPHI Collaboration reported the data on 
hadron production in
$e^{+}e^{-}$~ annihilations for the center of 
mass (c.m.) energy of $\sqrt{s} =
91$GeV finding ${\gamma}_2 = 0.04$~\cite{delphi}~. In 
hadron-hadron collisions
${\pi}^{+}-p$~, $K^{+}-p$~, $p-p$~, $p-{\bar p}$~ for 
c.m. energies ranging up
to 1000 GeV\cite{carr,hadron}~, values of ${\gamma}_2$~ 
increase from about 0.05 to
0.3 as energies increase to collider values. 
Distribution of galaxy counts in
the regions of sky covered by the Zwicky 
catalogue\cite{zwicky}~ yields
${\gamma_2} \simeq 0.3$~\cite{carr1}. Independently of 
the question whether the KNO scaling holds in all those 
different physical systems
, the measured values of ${\gamma}_2$~ clearly exclude 
the exponential region of Gaussian FIB process.
Much more information could be extracted if in addition to the 
moments of the
multiplicity distribution also the mass distribution 
would be available. In high energy lepton and/or hadron collisions for example
, this would require measuring the hadron mass distribution.

In conclusion, we have demonstrated that the 
off-equilibrium binary
fragmentation with scale-invariant fragmentation kernel 
and the scale-dependent
inactivation simulating the dissipation at small scales, 
yields the fragment mass and fragment multiplicity 
distributions 
which are scale-invariant
for a broad range of parameters. This is an important 
finding because most of
fragmentation processes in nature which have these 
scale-invariant features
are probably not associated with the dissipative 
processes acting at all
scales. The scale-dependent fragmentation processes 
may also develop
strong scale-invariant fluctuations (the KNO scaling) 
though the region of
their appearance is restricted to the particular value 
of exponent : $\alpha = -1$~, of the homogeneous 
fragmentation function. The
region at $\alpha = -1$~ and $\sigma$~ above $\sim 0.5$~ is the 
critical transition region 
of Gaussian FIB process. For 
other values of $\alpha$~, the fragment multiplicity 
distributions obey
the BPL scaling, i.e. the small amplitude  
limit of scaling multiplicity fluctuations. 
Another transition zone of the Gaussian FIB model is 
defined by the
width $\sigma$~ of inactivation rate function. At 
$\sigma \simeq 0.5$~, the
fragment size distribution changes from exponential (for 
$\sigma < 0.5$)~ into
power law (for $\sigma > 0.5$)~. 
The form of scaling function $\Phi(z_{\delta})$~, 
together with the form of  
fragment mass distribution $n(k)$~ 
impose strong constraints on the choice of basic 
functions of FIB
kinetic equations : the fragmentation and inactivation 
functions. This has
been demonstrated on the example of hadron 
production data in the $e^{+} e^{-}$~ 
annihilation\cite{zphys}~. Results of this Letter show that the closing of
gap between experimental observables related to the 
fragment mass distribution and/or the fragment multiplicity 
distribution and the basic ingredients of the kinetic 
theory, i.e. the rates of activation $F_{j,k-j}$~ and 
inactivation $I_k$~, can be achieved for many 
physical systems in the nature.

\vfill
 
\newpage
 
{\bf Figure captions}\\
 
{\bf Fig. 1}\\
Multiplicity probability distributions in the scaling 
variables (see eq. (\ref{b4})),
and the fragment mass distribution for two homogeneous 
fragmentation kernels 
and two Gaussian inactivation rate functions. Each set 
of data corresponds to 
$10^6$ independent events of Monte-Carlo simulations. \\
(i) Upper left part : the fragmentation kernel with 
$\alpha= -1$~ and the
inactivation rate function (\ref{inact}) for 
$c=1$ and two typical values of $\sigma$. 
Two sets of data are plotted for two different
total mass : $N=1024$ (crosses) and $N=4096$ (circles). 
These data are plotted
in the KNO form, i.e. : $\delta=1$ (see eq. 
(\ref{b4})).\\
(ii) Upper right part : the fragment mass distributions 
in a double-logarithmic 
scale are shown for the same parameters $\alpha , c, 
\sigma$~ as in (i). The total mass is $N=4096$. Big 
stars represent
results obtained for the same value of 
$\alpha , c$~ parameters and for a much larger value of 
$\sigma$~ 
($\sigma=10$~) , to show the independence of the 
scaling part of the fragment mass distribution with the 
value of $\sigma$. The line in between points is shown to guide the eyes.\\
(iii) Lower left part : the same as in (i) but for the 
fragmentation kernel 
with $\alpha=+1$~. These data are plotted
in the BPL form, i.e. : $\delta=1/2$~ (see eq. 
(\ref{b4})).\\
(iv) Lower right part : the fragment mass distributions 
for $\alpha=+1$. Parameters $c, \sigma , N$~ as in (ii).

{\bf Fig. 2}\\
The cumulant factorial moment ${\gamma}_2$~ of the 
fragment multiplicity distribution is
plotted vs the width parameter $\sigma$~ of the Gaussian 
inactivation function (\ref{inact}) with $c = 0.5, 1, 
5$~. The homogeneous 
fragmentation kernel is taken
with $\alpha = -1$~. Each point corresponds to system of 
size $N = 2^{18}$~,
and the values of ${\gamma}_2$~ are calculated by 
solving exact recurrent equations. The line joining points is shown to guide
the eyes.


\begin{references}
\bibitem{foot}
The factorial cumulant moments are 
integrals of corresponding cumulants which measure physical density
correlations free from trivial background of lower order density
correlations\cite{carr}~.
\end{references}
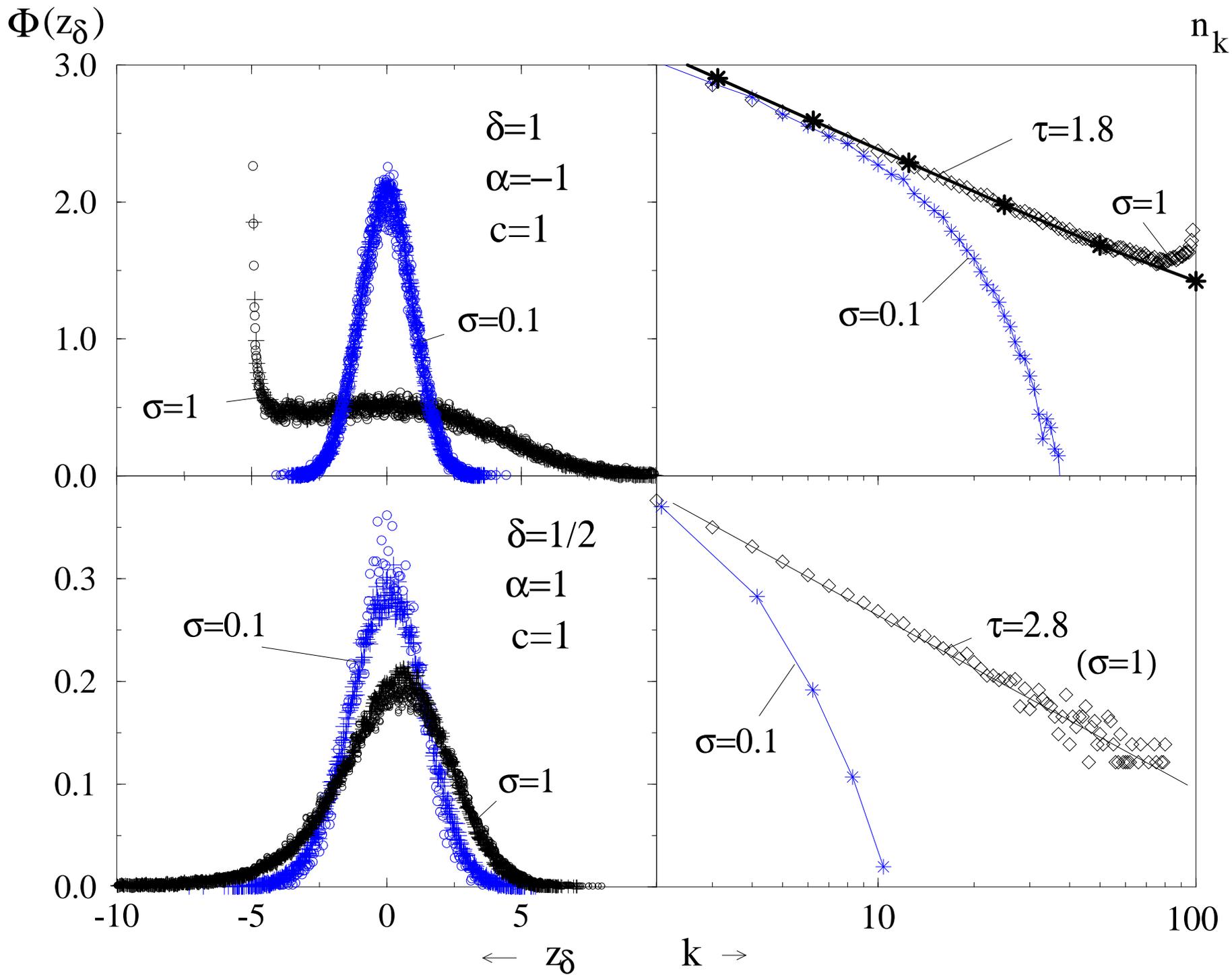
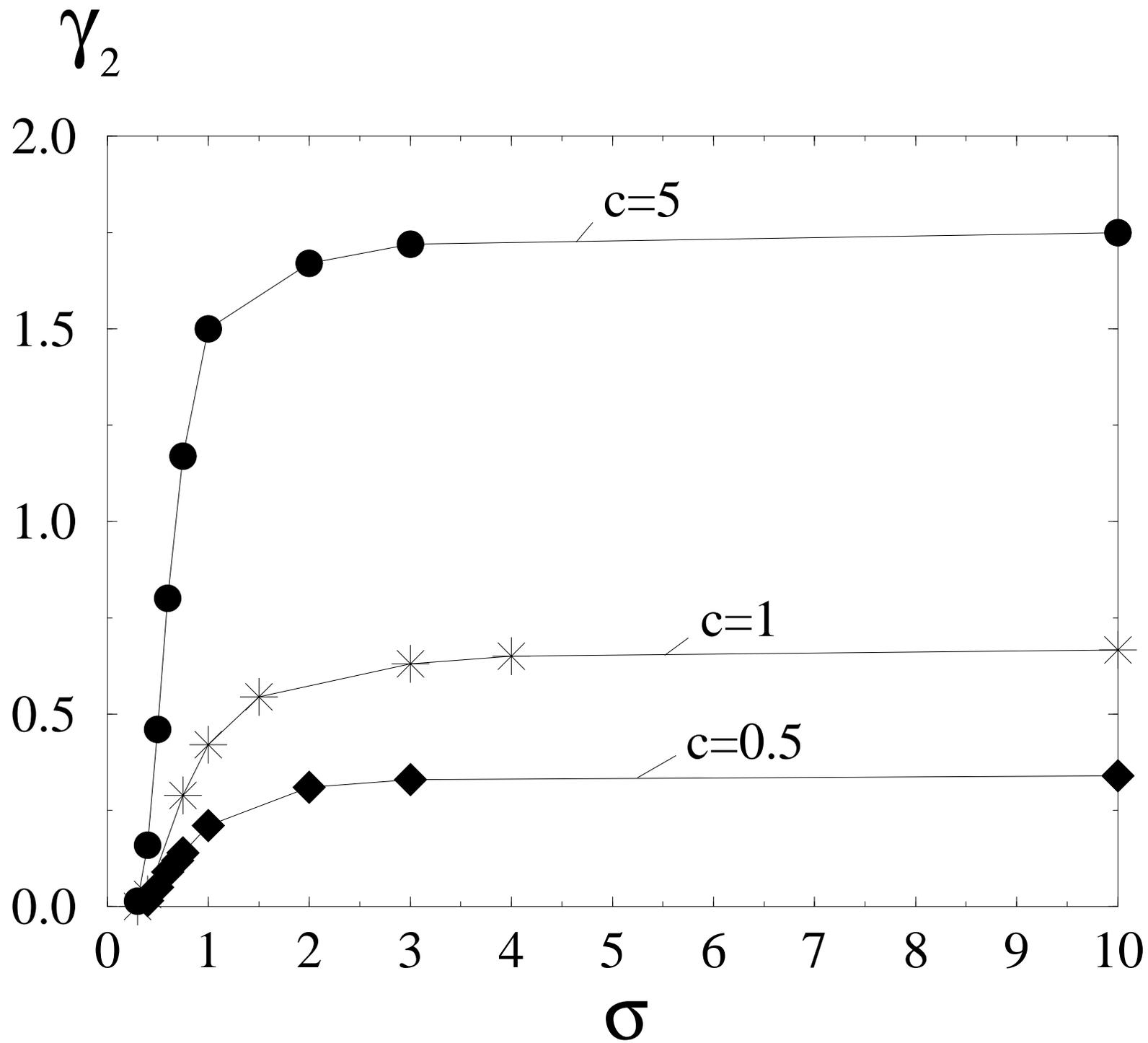
\end{document}